\documentclass[12pt,floatfix,pre]{revtex4}
\usepackage{subfigure}
\usepackage{soul}
\usepackage{graphicx}
\usepackage{dcolumn}
\usepackage{bm}
\usepackage{natbib}
\usepackage{amsmath, amsthm, amssymb}
\usepackage{color}
\usepackage{mathrsfs}
\usepackage{textcomp}

\begin{document}
\title{Bubble formation during the collision of a sessile drop with a meniscus}

\author{ D.L. Keij$^{1,2}$, K.G. Winkels$^1$, H. Castelijns$^2$, M. Riepen$^2$ and J.H. Snoeijer$^1$}
\affiliation{$^1$Physics of Fluids, Faculty of Science and Technology and Mesa+ Institute, University of Twente, The Netherlands\\
$^2$ASML BV, Veldhoven, The Netherlands}

\renewcommand{\today}{Dec, 2012}

\begin{abstract}
The impact of a sessile droplet with a moving meniscus, as encountered in processes such as dip-coating, generically leads to the entrapment of small air bubbles. Here we experimentally study this process of bubble formation by looking through the liquid using high-speed imaging. Our central finding is that the size of the entrapped bubble crucially depends on the location where coalescence between the drop and the moving meniscus is initiated: (i) at a finite height above the substrate, or (ii) exactly at the contact line. In the first case, we typically find bubble sizes of the order of a few microns, independent of the size and speed of the impacting drop. By contrast, the bubbles that are formed when coalescence starts at the contact line become increasingly large, as the size or the velocity of the impacting drop is increased. We show how these observations can be explained from a balance between the lubrication pressure in the air layer and the capillary pressure of the drop. 
\end{abstract}

\maketitle
\section{Introduction} \label{sec:chap4A}

Air bubbles can be entrapped in a wide variety of flows with a free surface \cite{Yarin:2006,Oguz:1990JFM, Tsai:2010, Tsai:2009Langmuir,Liow:2004,Liow:2007, Mason:1960,Deng:2007JFM,Thoroddsen:2003vs,Mesler:1986vs,vanDam:2004,Mandre:2009PRL, Mani:2010JFM,Driscoll:2010PRE,Hicks:2011, vanderVeen:2012ib,BillinghamKing}. Such entrapped bubbles can have large effects on phenomena such as bulk gas concentrations and sound emission \cite{Leighton:2004, Blanchard:1957te}. In addition, bubbles can be a nuisance for industrial processes, where they cause defects, obstructions or noise \cite{Switkes:2005kv}. Various different mechanisms for air entrapment have been proposed in the literature. For example, it is well known that the impact of a drop on a free surface (e.g. rain) can induce an inertial cavity collapse that results in an oscillating bubble \cite{Oguz:1990JFM, Deng:2007JFM}. At lower impact speeds, air films can be trapped between a liquid drop and a liquid or a solid wall, by lubricating effects of the medium. Such a lubricating film can delay or completely avoid coalescence. In the extreme case such an air film can even cause a droplet to float on the liquid surface \cite{Couder:2005Nature,Couder:2005PRL,Gilet:2009JFM}. Depending on whether this film simply drains or becomes unstable, air entrainment might occur \cite{Thoroddsen:2003vs,Liow:2007,Mesler:1986vs, Sigler:1990}. However, it has been predicted that bubbles can also be entrapped during a coalescence with zero impact velocity, for which no lubricating film develops. In this case the bubbles form by reconnecting capillary waves that result in toroidal bubble rings as described theoretically \cite{Duchemin:2003JFM}. 

\begin{figure}
	\centering
	\includegraphics[width=0.99 \textwidth]{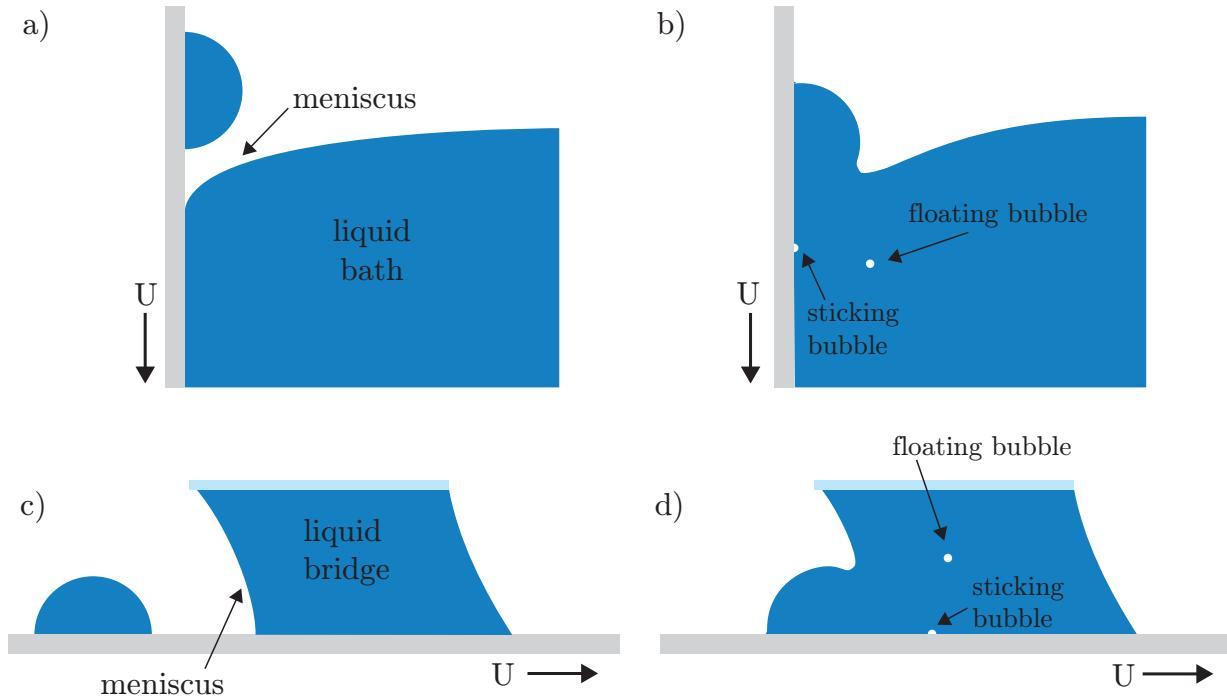}
 	\caption{Two examples of bubble entrapment during the collision of a sessile drop with a meniscus: dip-coating (a,b) and immersion lithography (c,d). (a) A sessile drop moves with the substrate into a liquid reservoir. (b) Air bubbles can be trapped during the coalescence of the drop with the reservoir. These bubbles are sometimes floating in the bulk liquid (floating bubbles) and sometimes attached to the substrate (sticking bubbles). (c) A sessile drop moves with the substrate into a liquid bridge. The liquid bridge consists of a gap filled with liquid, stationary at the top (pinned contact line) and sliding over the substrate at the bottom (moving contact line). The front part of the liquid bridge forms the meniscus in the 'drop-meniscus-collision'. (d) Similar as in dip-coating, two types of bubbles can be formed. }
 	  \label{fig:chap4_problemsketch}
\end{figure}

In this paper we focus on entrapment of air bubbles that form when a sessile drop impacts with a moving meniscus. This is relevant for applications such as dip-coating and immersion lithography (Fig.~\ref{fig:chap4_problemsketch}). Dip-coating is a very common setup both for applications and fundamental studies, where a solid plate is plunged into or withdrawn from a liquid reservoir. A very similar geometry is present in immersion lithography, a technology used in semi-conductor industry: replacing the air in between a lens and the silicon wafer by a liquid leads to an increase in the numerical aperture of the system, allowing for the projection of smaller structures. A simplified version of the flow geometry for immersion lithography is sketched in  figure~\ref{fig:chap4_problemsketch}~c-d). The ``meniscus'' consists of a liquid bridge held between the hydrophilic glass plate and a wafer which is usually made partially wetting by a coating. The water will preferably remain in the gap, due to the contact with the hydrophilic glass plate, even when the substrate is in motion. A first mechanism for bubble entrapment is that at high velocities, the contact line can become unstable resulting in entrainment of a thin air film \cite{Chan:2012} or bubbles \cite{Benkreira:2008,Benkreira:2010}. A second  mechanism that leads to bubbles in these applications is due to residual drops, which are left on the substrate. These drops move along with the substrate and collide with the reservoir (dip-coating) or liquid bridge (immersion lithography). Air bubbles are generically entrapped during a collision of such a sessile drop with the meniscus.

\begin{figure}
	\centering
	\includegraphics[width=0.7 \textwidth]{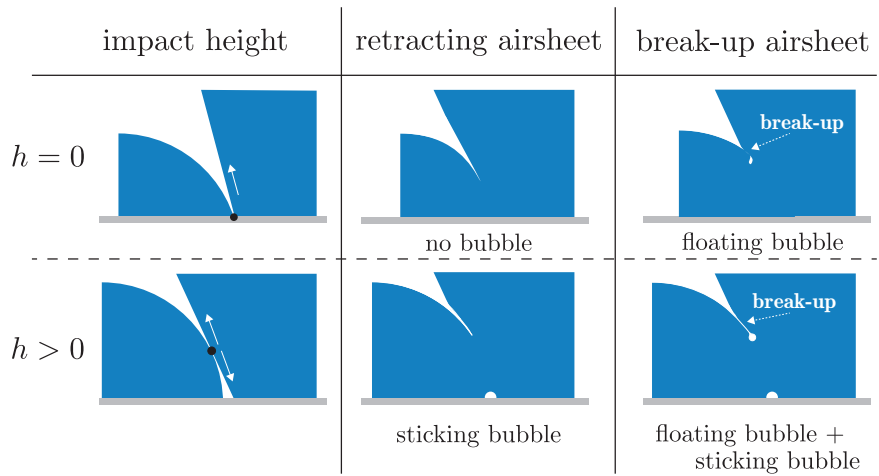}
	\caption{Scenarios for bubble entrapment. Two different bubble types can form depending on the impact height $h$ (indicated by the black dot). If the first contact is located at the contact line ($h=0$), the upward motion of the coalescing bridge drives the air between the drop and the meniscus away from the substrate. Depending on the dynamics of this retraction, this results into breakup (floating bubble) or no bubble. Similar dynamics is observed when first contact is above the contact line $h>0$. However, for the downward motion of the liquid bridge we always observe the entrapment of an air bubble at the substrate (sticking bubble).}
	  \label{fig:chap4_overview}
\end{figure}

We present an experimental study on the formation of bubbles resulting from the impact of a sessile drop with a meniscus close to a moving contact line. The sessile drop always coalesces with the meniscus. As summarized in Fig.~\ref{fig:chap4_overview}, however, we identified two scenarios for bubble entrapment. First, when the contact takes place at or very close to the moving contact line (i.e. at $h=0$, with $h$ the impact height with respect to the moving substrate) we observe ``floating bubbles''. These bubbles are spherical and float in the bulk of the liquid bridge. Second, when the first contact between the sessile drop and the meniscus takes place above the moving contact line ($h>0$) also a ``sticking bubble'' can be formed. This bubble is attached to the substrate and moves with the wafer. It should be noted that apart from these two cases, the coalescence can also result into no bubble formation, or a combination of floating bubbles and sticking bubbles. As summarized in Fig.~\ref{fig:chap4_overview}, this depends on whether or not the airsheet breaks up during the coalescence.

The paper is organized as follows. We first introduce the experimental setup (Sec.~\ref{sec:chap4_setup}). Then we will discuss the formation mechanism of the two possible types of bubbles appearing during the drop-meniscus collision (Sec~\ref{sec:chap4observations}), and quantify the size of the entrapped bubbles, as a function of the impact velocity and the size of the sessile drop (Sec.~\ref{sec:chap4bubblesizes}). The latter section also includes explanations for the size of the two types of bubbles. The paper ends with a discussion (section~\ref{sec:chap4_discussion}).

\section{Experimental setup} \label{sec:chap4_setup}

\begin{figure}
	\centering 	
	\includegraphics[width=0.99 \textwidth]{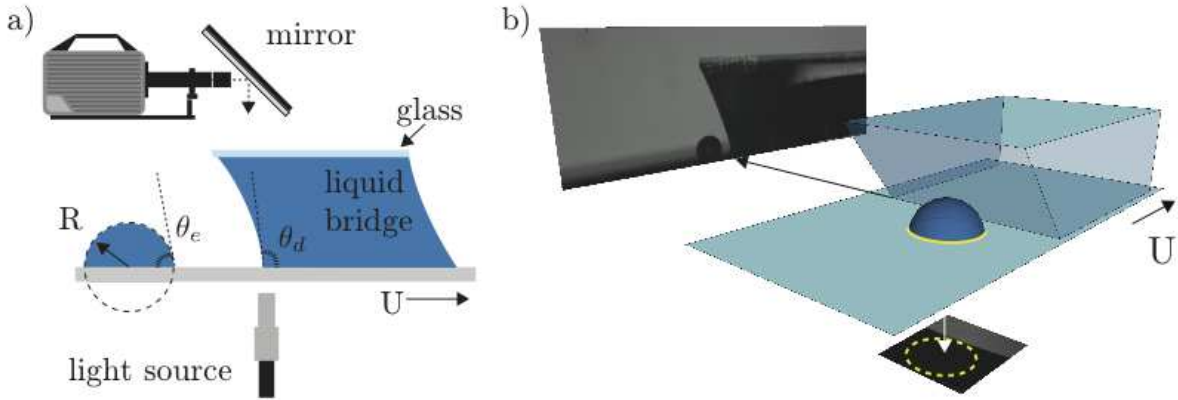}
	\caption{(a) Schematic of the experimental setup. The top-view recordings are obtained with backlit illumination. The drop radius $R$, equilibrium contact angle $\theta_{e}$ and dynamic contact angle $\theta_{d}$ are also defined. (b) Three-dimensional impression of the impact of a sessile droplet (yellow line denotes the contact line) with the liquid bridge. The black-and-white images show typical recordings of the side- and top-view obtained with the experimental setup.} 
	\label{fig:chap4_setup}
\end{figure}

The experimental setup consists of a coated glass wafer ($D = 300$ mm) clamped to a turntable that is rotated with controlled angular velocity $\omega = 5 - 225$ $^\circ /s$. The rotational motion of the wafer approximates a linear motion in the camera reference frame, due to the small droplet size compared to the radial position on the wafer  ($r \sim 135$ mm). A detailed description of a similar setup is given in Winkels et al. \cite{Winkels:2011}. The geometry in which the collision between a sessile drop and a meniscus is realized is sketched in Fig.~\ref{fig:chap4_setup}. A small glass plate ($7 \times 12$ mm) is fixed in the camera reference frame at a height $0.8-2.8$ mm above the substrate and close ($<15$ mm) to the edge of the wafer. The gap between the wafer and the glass plate is filled with water (Millipore, Milli-Q, Advantage A10) resulting in a liquid bridge with a pinned contact line at the hydrophilic glass plate and a mobile contact line at the coated wafer ($\theta_e \sim 70^\circ - 83^\circ$). With this construction, the liquid is held fixed in the camera reference frame also when the wafer is rotated. The liquid bridge then slides over the wafer, such that at the front and at the rear of the bridge there is a moving contact line. Within the experimental range of velocities, we observe no air entrainment unless there is a collision with an incoming sessile drop. 

The two main control parameters in the experiment are the size and velocity of the impacting drop. Before each measurement a sessile droplet is created on the wafer with multiple drops from a micro-drop generator (Microdrop Technologies, MD-E-3000 in combination with a MD-K-130 dispenser head, single drop diameter $\sim 65$ $\mu m$). We achieve drop sizes in the range $R \sim 0.2-1$ mm. The sizes are determined from side view recordings. The wafer velocity, and hence the impact velocity of the sessile drop, is varied over the range $U \sim 0.01-0.54$ m/s.

 The process of impact and the subsequent bubble formation is recorded simultaneously with two high speed cameras. For the top view a Shimadzu HPV1 or Photron SA1.1 (framerates $10-250$ kfps) is used, connected to a long distance microscope (Navitar 12X Zoom with 1.5X front lens) obtaining a maximum resolving power of 2 $\mu \text{m/pixel}$. The side view is captured simultaneously with a PCO1200s camera (used at $1$ kfps) attached to a lens (Jenoptik, JENmetar 1x/12 LD). To image the formation of the bubbles, the meniscus is viewed from above through the top glass plate, as shown in Fig.~\ref{fig:chap4_setup}a. In combination with backlit illumination the meniscus turns black, while the wetted area inside the liquid bridge turns bright. Typical image recordings are shown in Fig.~\ref{fig:chap4_setup}b. The contact line of the sessile drop is marked with the yellow line: prior to impact, the drop is 'hidden' behind the meniscus in the top view recordings. During the coalescence, the region where the drop and meniscus have merged will actually turn bright and can thus be monitored very accurately. Note that the white ring that appears in the bright part of the image is an optical artefact without fluid mechanical meaning (see Fig.~\ref{fig:chap4_floating}a).
 
 Finally, we note that the liquid bridge can exhibit a rather complex geometry. The meniscus can be convex or concave depending on the gap height, volume of liquid and the dynamic contact angle. Throughout our experiments we have tuned these parameters in order to keep the impacting meniscus as ``flat'' as possible in the side view. By avoiding strongly concave or convex menisci, we have been able to obtain conditions where the bubble formation is highly reproducible. Namely, the geometry of impact is then a spherical cap with angle $\theta_e$, colliding with a meniscus that can be approximated by a ``plane'' with dynamic contact angle $\theta_d$. Then a simple geometrical argument suggests that if $\theta_e + \theta_d < \pi$, $h = 0$, while if $\theta_e + \theta_d > \pi$, $h > 0$. The combination with for example the dynamic contact angle as a function of velocity, would give a possible predictive tool for the impact height $h$ as a function of velocity -- provided that meniscus profile can be kept perfectly flat. In our experiments we observe the impact height from the top view measurements and investigate the resulting coalescence dynamics and possible bubble formation process.
 
\section{Experimental observations}\label{sec:chap4observations}

\subsection{Floating bubbles: $h=0$} \label{subsec:chap4ER_floating}

\begin{figure}
	\centering 	
	\includegraphics[width=0.75 \textwidth]{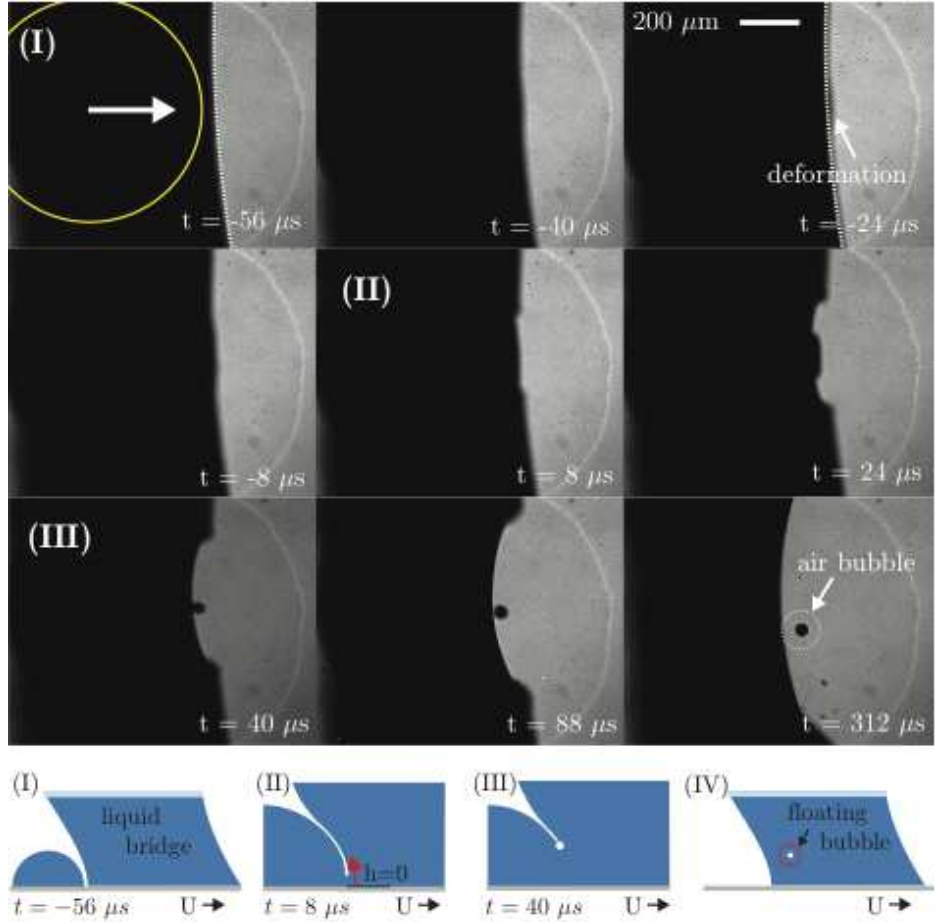}
	\caption{Top: High-speed recordings of the formation of a floating bubble. The yellow solid line denotes the contact line of the sessile droplet before impact (hidden behind the black meniscus). In the final top view image ($t=312 \mu$s), the air bubble is located in the white dashed circle. Parameter settings: $U = 300 \pm 3$ mm/s, $R =  0.35 \pm 0.02$ mm, $R_{bub} =22 \pm 3$  $\mu m$. Note: the white ring in the bright part of the image is an optical artefact. Bottom: Sketch of the side view during the bubble formation at different moments in time. I) The sessile drop approaches the meniscus at velocity $U$. II) Before coalescence an air sheet is formed between the meniscus and the drop, which retracts upwards away from the substrate if coalescence is initiated at the contact line  ($h=0$). III) The retracting air sheet  is enclosed during the coalescence. IV) A spherical bubble is formed, that floats in the bulk liquid (floating bubble).} 
	\label{fig:chap4_floating}
\end{figure}

We first consider the case where the coalescence is initiated at the contact line, with impact height $h=0$ within experimental accuracy. The subsequent coalescence is  directed upwards, away from the substrate. The outcome of the experiment is that either a floating bubble is formed inside the liquid, or no bubble is formed. Figure~\ref{fig:chap4_floating} shows a typical image sequence in the case a floating bubble is entrapped. Below the experimental stills we provide sketches of the side view, to clarify the various stages of the bubble formation. These sketches serve as illustration and do not provide a fully accurate description of the deformation.

A sessile droplet with size $R$, moves from left to right with velocity $U$ (Fig.\ref{fig:chap4_floating}I). The yellow solid line indicates the contact line of the sessile drop which is hidden behind the black meniscus prior to impact. The white dashed line denotes the moving contact line of the meniscus. Interestingly, a small deformation of the contact line can be observed at time $t= -24$ $\mu s$ before impact (white arrow, Fig.~\ref{fig:chap4_floating}, $t = -24$ $\mu s$). This deformation is due to the lubrication pressure building up in the air between the drop and the meniscus. Subsequently, the coalescence process starts at, or at least very close to, the contact line (Fig.~\ref{fig:chap4_floating}II). During the upward motion of the coalescing bridge a pocket of air is enclosed, resulting into a floating bubble (Fig.~\ref{fig:chap4_floating}III). The coalescence continues while the spherical bubble floats inside the liquid bridge ($t=312$ $\mu s$, Fig.\ref{fig:chap4_floating}). The floating bubble is stable and remains inside the liquid bridge also after drop collision (Fig.~\ref{fig:chap4_floating}IV). 

\subsection{Sticking bubbles: $h>0$} \label{subsec:chap4ER_cl}

\begin{figure}
	\centering 	
	\includegraphics[width=0.75 \textwidth]{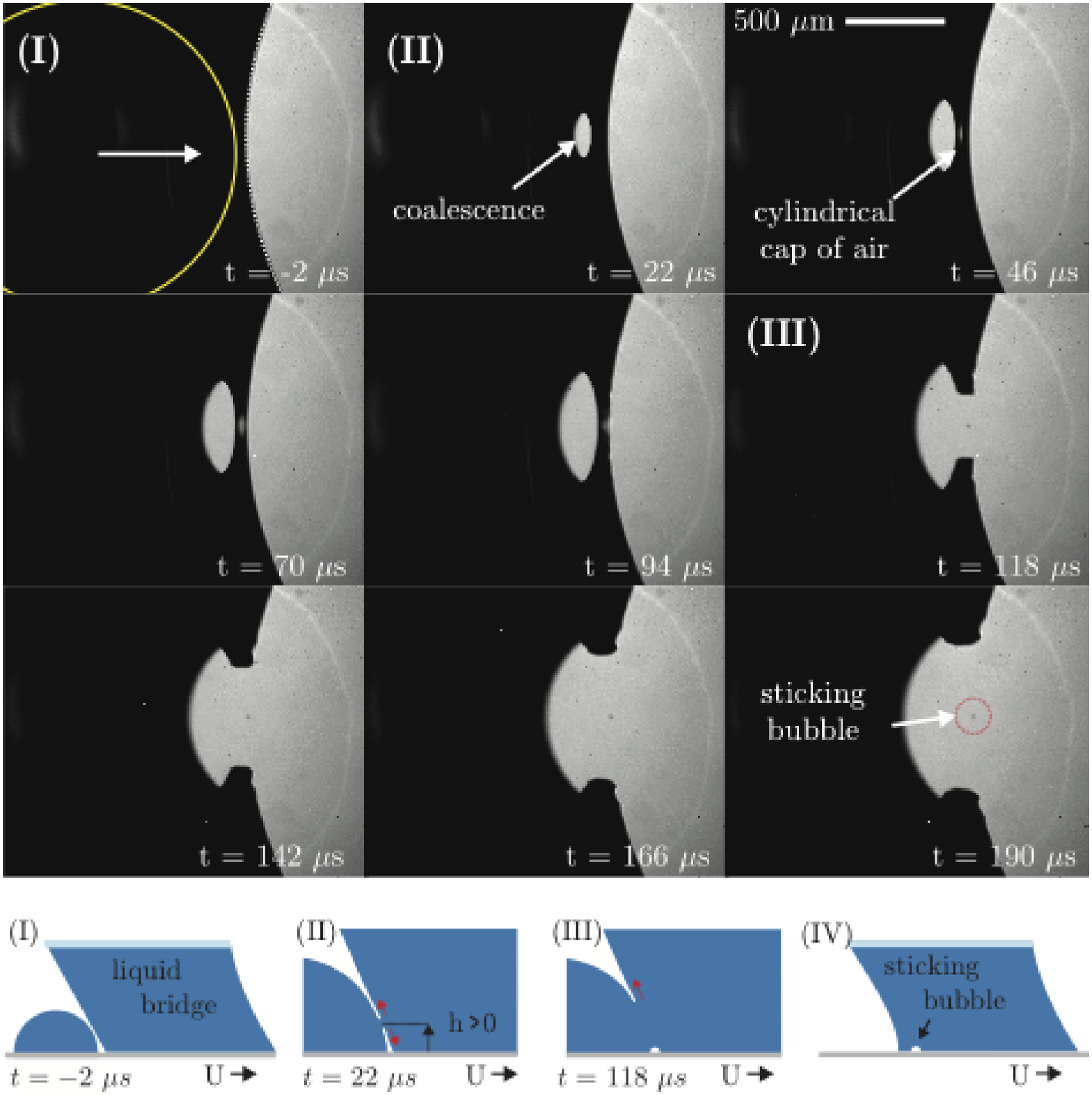}
	\caption{Top: High-speed recordings of the formation of a sticking bubble. The yellow solid line in the first image denotes the static contact line of the sessile droplet (hidden behind the meniscus), while the dashed line shows the moving contact line of the meniscus. In the final image ($t = \text{190}\mu s$), the air bubble is indicated by the red dashed circle. Parameter settings: $U = 375 \pm 3$ mm/s, $R =1.00 \pm 0.03$ mm, $R_{bubble} = 12\pm 6$ $\mu m$. Bottom: Sketch of the side view during the bubble formation. I) The sessile drop approaches the meniscus. II) First contact occurs at finite height $h>0$. As a result, a coalescing bridge moves upwards and downwards. III) The downward retracting air sheet creates an air channel at the substrate (perpendicular to the field of view), which finally breaks up into a sticking bubble.  IV) The sessile drop merged with the liquid bridge, leaving a sticking bubble inside the bulk liquid.} 
	\label{fig:chap4_cl}
\end{figure}	

Another type of bubble that can be entrapped is a sticking bubble, attached to the substrate after its formation. These sticking bubbles form a spherical cap with a finite contact angle, and arise when the initial contact between the drop and the substrate occurs at a finite height above the substrate, $h>0$. The process of sticking bubble formation is clearly revealed in the top view recordings as shown in Figure~\ref{fig:chap4_cl}. Once more, we complement the still images with side view sketches to clarify the process of bubble formation (I-IV). It should be noted that in comparison with Fig.~\ref{fig:chap4_floating}, besides the impact height $h>0$ also the drop size $R$ is different. However as will be shown later this does not affect the phenomenon.

At very short time before impact ($t = -2$ $\mu$s, Fig. \ref{fig:chap4_cl}I), the drop-meniscus separation distance is very small and the sessile drop is indicated by the yellow circle. Once again, the meniscus is completely black and the drop moves in from the left with velocity $U$, hidden behind the meniscus in this top view. The moving contact line of the meniscus is slightly curved and indicated by the white dashed line. After several microseconds, $t = 22$ $\mu s$, one clearly observes that contact has occurred at a finite height above the substrate. The bridge connecting the drop and the meniscus appears as a bright ellipsoidal area (Fig. \ref{fig:chap4_cl}II). Coalescence is initiated in all directions, and the downward part of the bridge approaches the substrate (Fig. \ref{fig:chap4_cl}, $t= 22-94$ $\mu s$). During this rapid motion, the air is confined in a cylindrical cap that is squeezed between the wafer and the downward coalescing front. At the final stage of bubble formation, this cylindrical shape pinches off symmetrically along the wafer (up- and downwards in the image), leaving a small ``satellite'' bubble at the substrate. This sticking bubble (Fig. \ref{fig:chap4_cl}III, $ t = 190$ $\mu s$, red dashed circle), moves with the substrate. The coalescence also proceeds in the upward direction, away from the contact line. In some experiments we have seen that this can also result into a floating bubble, in analogy to those described in Sec.~\ref{subsec:chap4ER_floating}.  

\section{Bubble sizes} \label{sec:chap4bubblesizes}

\subsection{Measurements}\label{quantitative_results}

\begin{figure}
	\centering 	
	\includegraphics[width=0.6 \textwidth]{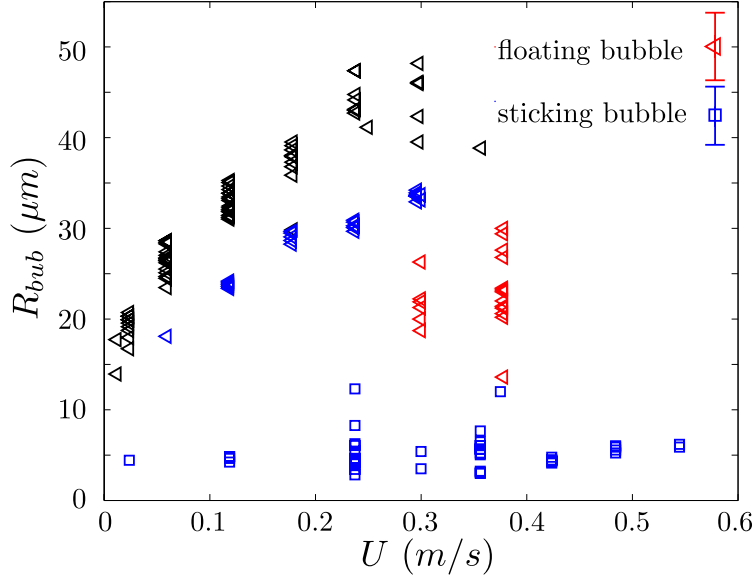}
	\caption{Size of entrapped bubbles $R_{bub}$, measured as a function of impact velocity $U$. The data summarize all collision events, for varying drop sizes. The triangles ($\triangleleft$) represent the data for floating bubbles. For these bubbles the colors indicate the approximate impact drop radius: red triangles (\textcolor{red}{$\triangleleft$}) $R<400$ $\mu m$, blue triangles (\textcolor{blue}{$\triangleleft$}) $400<R<600$ $\mu m$ and black triangles (\textcolor{black}{$\triangleleft$}) $R>600$ $\mu m$. The squares (\textcolor{blue}{$\square$}) denote the sticking bubbles for $0.21<R<0.62$ mm. The typical measurement error is indicated in the legend.} 
	\label{fig:chap4_expdata}
\end{figure}

We now investigate the size of the entrapped bubbles, characterized by the radius $R_{bub}$, for different impact conditions. In particular, we vary the size of the sessile drop, $R$, and the impact velocity, $U$. Our measurements are summarized in Fig.~\ref{fig:chap4_expdata}. The triangles ($\triangleleft$,\textcolor{red}{$\triangleleft$},\textcolor{blue}{$\triangleleft$}) show data obtained for floating bubbles, for $U=0.01$ - $0.4$ m/s and $R=0.26$-$1.00$ mm in radius. The blue squares (\textcolor{blue}{$\square$}) are results for sticking bubbles, for the impact velocities $U = 0.02 - 0.54$ m/s and $R = 0.21-0.62$  mm. Each data point corresponds to a single collision, and the error bars represent the accuracy of resolution.

One clearly observes that the two different types of bubbles display very different sizes, originating from the different mechanisms of their formation. For floating bubbles, the general trend is that $R_{bub}$ increases with impact velocity. However, for a given velocity there is a rather large spread of the data. We show below that this can be attributed to the different sizes of impacting drops. This is also why $R_{bub}$ is lower for the largest impact velocities, which were obtained for relatively small drops. The results for sticking bubbles are very different:  within our experimental resolution, the bubble size is independent of both the impact velocity and of the radius of impacting drop. The size of sticking bubbles is approximately $R_{bub}\sim 5~\mu$m, which is  much smaller than the floating bubbles.

\subsection{Floating bubbles: $h=0$} \label{chap4A_floating}

As observed in Sec.~\ref{subsec:chap4ER_floating}, the meniscus is weakly distorted already before the coalescence starts (arrow Fig.~\ref{fig:chap4_floating}, $t = -24$ $\mu s$). This suggests that a pressure builds up inside the air layer in between the drop and meniscus, which is sufficiently strong to deform the liquid interfaces prior to impact. In fact, this effect is well-known for drop impact on a solid or on a liquid reservoir \cite{Couder:2005PRL, Couder:2005Nature,Gilet:2009JFM,Mandre:2009PRL,Kolinski:2011,vanderVeen:2012ib, deRuiter:2012PRL}. As the air is squeezed out of the narrow gap, the viscous air flow leads to a lubrication pressure inside the layer. 

\begin{figure}[h!]
\centering
\includegraphics[width = 0.6 \textwidth]{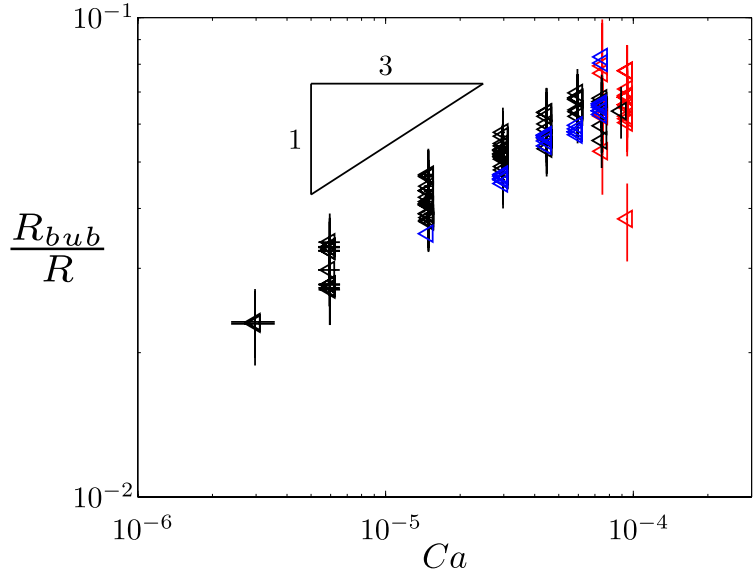}
\caption{Floating bubbles sizes. Bubble radius $R_{bub}$ normalized by $R$, plotted versus the capillary number ${\rm Ca}=U\eta_g/\gamma$.  The colors indicate the approximate impact drop radius: in red $R<400$ $\mu m$, in blue $400<R<600$ $\mu m$ and in black $R>600$ $\mu m$.}
\label{fig:chap4_floatingratio}
\end{figure}

The volume of the entrapped air sheet ultimately determines the size of the air bubble. We therefore propose that the lubrication pressure is responsible for the velocity dependence of $R_{bub}$. As the pressure originates from the dynamical viscosity of the gas, $\eta_g$, it is natural to re-plot the experimental data in terms of the capillary number ${\rm Ca}=U\eta_g/\gamma$, where $\gamma$ is the surface tension (even though we did not explicitly vary the gas viscosity $\eta_g$ and surface tension $\gamma$). Figure~\ref{fig:chap4_floatingratio} shows the data in dimensionless form, where we scaled the bubble radius by the radius of the impacting drop. In comparison to Fig.~\ref{fig:chap4_expdata}, we indeed observe that the rescaled bubble sizes are nicely aligned, and display a monotonic increase with ${\rm Ca}$. The scatter is comparable to the experimental error in determining the bubble radius. The data are reasonably described by a power-law over more than a decade in ${\rm Ca}$. 

\begin{figure}[h!]
	\includegraphics[width=0.5 \textwidth]{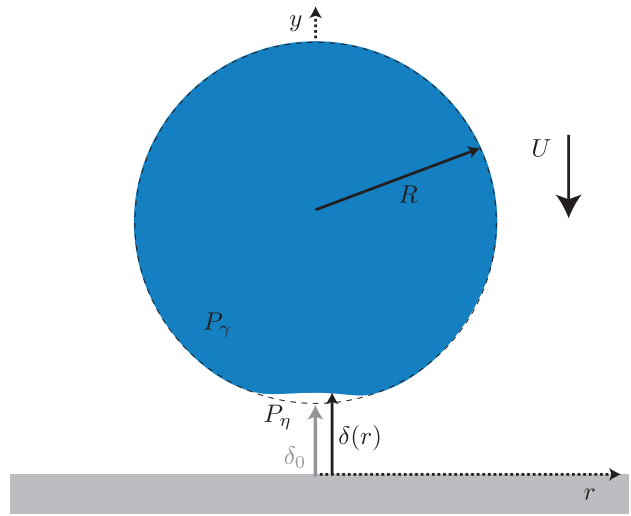}
	\caption{Sketch of a drop impacting a substrate.} 
	\label{fig:fig_chap4_dropDeformation}
\end{figure}

We now rationalize this dependence on ${\rm Ca}$ by drawing the analogy with drop impact on a solid substrate \cite{Mandre:2009PRL,Mani:2010JFM,Bouwhuis:2012ARXIV}. The geometry of this impact is sketched in Fig.~\ref{fig:fig_chap4_dropDeformation}, and is actually much simpler than the geometry of Fig.~\ref{fig:chap4_problemsketch}. While the drop approaches the substrate with velocity $U$, the thickness of the air layer reduces and air is squeezed out. As long as the lubrication pressure, $P_{\eta}$, is significantly smaller than the capillary pressure inside the drop, $P_{\gamma} = 2 \gamma/R$, the drop remains spherical. In this regime the drop is undeformed and the shape of the air layer is described by $\delta(r) \simeq \delta_0 +r^2/(2R)$. Then, the lubrication pressure can be computed exactly \cite{Hinch:1986JFM}.

\begin{equation}
P_\eta = \frac{3 \eta_{g} U R}{{\delta_0}^2 \left(1 + \frac{r^2}{2 R \delta_0}\right)^2},
\end{equation}
where $\delta_0$ is the gap thickness at $r=0$. From this equation, it is clear that the typical lateral length scale of the lubricating pressure is, $L \sim (R \delta_0)^{1/2}$. 

To estimate at what distance $\delta_0$ the drop will start to deform, we balance $P_{\eta} \sim P_{\gamma}$. Note that this is different from \cite{Mandre:2009PRL,Mani:2010JFM}, in which $P_\eta$ is balanced with the inertial pressure inside the liquid. In the experiments presented in this paper the impact velocities and drop impact size are smaller than \cite{Mandre:2009PRL,Mani:2010JFM}, such that the inertial pressure is in fact sub-dominant compared to the capillary pressure. For the smallest drop size and lowest velocities, it is relevant to question whether viscous effects inside the drop come into play~\cite{KL09,BHL07}. However, an estimation of the smallest Reynolds number, Re $\sim 2$, shows that viscous effects in the liquid are comparable to inertia, and therefore are sub-dominant as well. The visco-capillary balance yields a characteristic height for the dimple below the drop $H$ (Fig.~\ref{fig:fig_chap4_dropDeformation}):

\begin{equation}
\frac{3 \eta_{g} U R}{H^2} \sim \frac{2 \gamma}{R} \Longrightarrow H \sim R \left( {\rm Ca}\right)^{1/2}.
\end{equation}
This determines the distance at which the drop first deforms, and provides the natural length scale for the dimple size in the direction normal to the wall. In a very short time ($\sim 0.1$ ms) the deformation finally leads to very fast coalescence during which the lubricating air sheet is trapped and forms a bubble. The volume of the air pocket, and hence of the bubble, is expected to scale as  

\begin{equation}
R_{bub}^3 \sim H L^2 \sim H^2 R \Longrightarrow \frac{R_{bub}}{R} \sim {\rm Ca}^{1/3}.
\end{equation}

Despite the obvious difference in impact geometry, this scaling argument is consistent with the experimental data for the floating bubbles that form during the impact of a sessile drop with a meniscus. This suggests that the bubble volume is indeed governed by the lubrication effect of the squeezed gas layer. Clearly, a much more detailed modelling of the impact geometry, and the subsequent drainage of air, is required to confirm this scenario. 

\subsection{Sticking bubbles: $h>0$}\label{chap4A_cl}

\begin{figure}
	\centering 	
	\includegraphics[width=0.5 \textwidth]{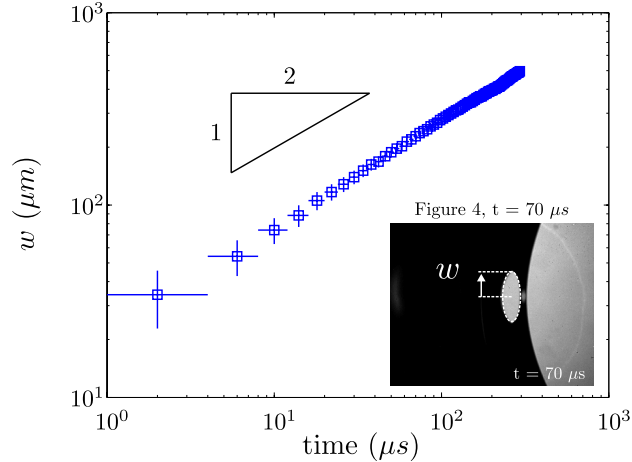}
	\caption{Coalescence bridge growth $w$ as a function of time $t$, for impact occurring at finite height $h$. Inset: typical recording image (see Fig.~\ref{fig:chap4_cl}, $t = 70$ $\mu s$) with definition of $w$.} 	\label{fig:fig_chap4_coalescence}
\end{figure}

How can we understand that the size of sticking bubbles is independent of $U$? The key is that the coalescence of the meniscus and drop in Fig.~\ref{fig:chap4_cl}, leading to the air confinement at the substrate, is much faster than the impact velocity $U$. This is quantified in Fig. \ref{fig:fig_chap4_coalescence}, where we plot the growth of the coalescing bridge in time. The bridge radius $w$ approximately grows as $w \sim t^{1/2}$ (best fit: $0.55$), which is comparable to the standard inertia-dominated coalescence of two free droplets \cite{Duchemin:2003JFM,Wu:2004}. The fact that the exponent is slightly larger than 1/2 is consistent with recent measurements on water drops, signalling a crossover to a viscosity-dominated coalescence \cite{Paulsen:2011}. A power-law with exponent smaller than unity implies a very rapid dynamics at the initial stages. From the perspective of bubble entrapment, this means that speed with which the cylindrical cap of air is squeezed (Fig.~\ref{fig:chap4_cl}), is much faster than the actual impact velocity of the sessile drop. This explains why $U$ is irrelevant for the size of the sticking bubbles.

\begin{figure}[h!]
	\centering 	
	\includegraphics[width=0.99 \textwidth]{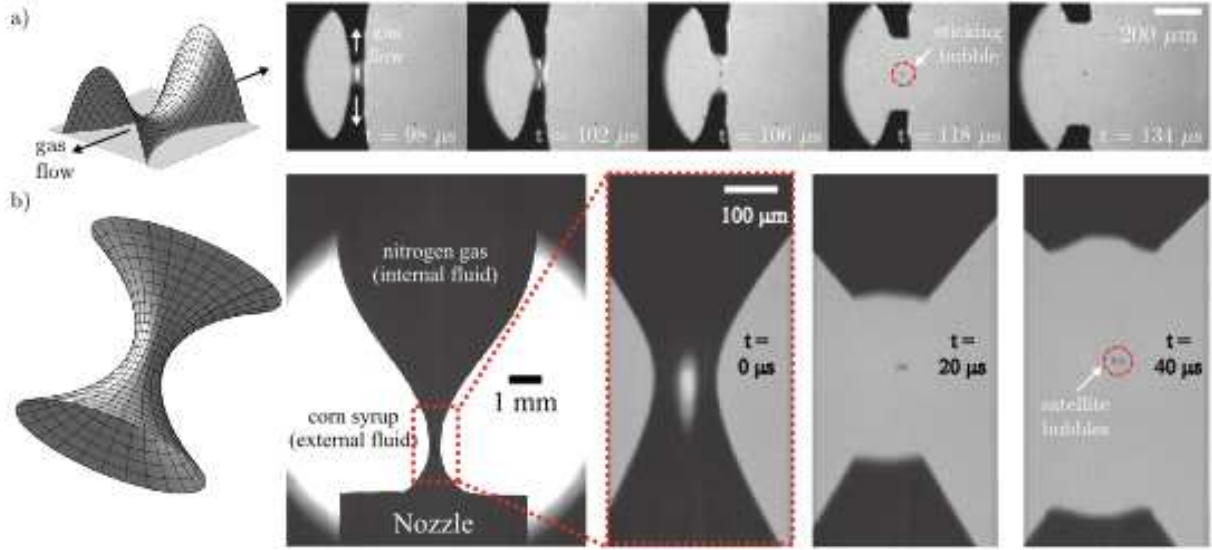}
	\caption{Analogy of break-up of a cylindrical cap (a) (zoom in top view Fig. ~\ref{fig:chap4_cl}) with the break-up of a cylindrical neck in bubble pinch-off (b) (reprint Burton et al. PRL 94, 184502, 2005). The first image per row is a three dimensional sketch of the liquid gas interface in each configuration. The break-up of a cylindrical cap, results in a sticking bubble, while the axi-symmetric neck in case of the buoyant bubble results in spherical satellite bubbles.}
	\label{fig:chap4_pinchoff}
\end{figure}

It still remains a question why the size of sticking bubbles is also independent of the drop radius $R$. A closer look at the sticking bubble formation, reveals a striking analogy with the pinch-off process of a buoyant air bubble from a nozzle in water, such as visualized by Burton et al~\cite{Burton:2005PRL} and shown in a time sequence in Fig.~\ref{fig:chap4_pinchoff}b. The first image per row is a three dimensional sketch of the liquid gas interface in each configuration. Apart from the presence of the substrate, the geometry of the liquid-gas meniscus is very similar in case of the sticking bubble formation. For the buoyant bubble, the detachment of the large bubble occurs by the breakup of a slender cylindrical neck. During this breakup, small satellite bubbles of approximately $10$ microns diameter are formed at the middle of the neck. According to \cite{Gordillo:2007PRL},  the size of these satellites can be attributed to the inertia in the gas phase, which becomes relevant during the final stages of pinch-off. Indeed, we observe a very similar breakup of the cylindrical cap in figure~\ref{fig:chap4_pinchoff}a. The main difference with respect to Burton et al. \cite{Burton:2005PRL} is that in our case the cylinder of air is attached to the surface. However, the time scale for breakup and the size of the satellite bubble are almost the same in both experiments. Clearly, the size of the impacting drop is completely irrelevant during this final stage of the bubble formation.

In conclusion, the sticking bubbles display a universal size due to a combination of two classical singularities: coalescence and breakup. First, the growth of the coalescing bridge is extremely fast, making the impact velocity of the sessile drop irrelevant. It confines the air between the coalescing bridge and the substrate into a cylindrical air pocket. This cylinder of air undergoes a pinch-off, similar to \cite{Burton:2005PRL,Eggers:2007PRL,Gordillo:2007PRL,Bergmann:2006PRL}. What remains is a small bubble that is completely independent of impact conditions.

\section{Discussion} \label{sec:chap4_discussion} 

We have identified two types of bubbles that form during the impact of a sessile drop with a meniscus. As summarized in Fig.~\ref{fig:chap4_overview}, the formation crucially depends on the height of the initial contact. When the first contact occurs at $h=0$ the resulting bubbles are ``floating'' inside the liquid. These bubbles appear due to the entrapment of a lubricating air film, in analogy for bubble entrapment below drops falling on a solid surface. The bubble sizes increase in size with increasing impact velocity, consistent with a scaling argument $R_{bub}/R \sim {\rm Ca}^{1/3}$. By contrast at finite impact height, $h>0$, the entrapped bubbles are ``sticking'' to the substrate. Their size is a few microns and their formation is completely independent of the size and velocity of the impacting drops. The reason for this is that the formation process is dominated by the very fast coalescence, followed by the very fast pinchoff of an air cylinder. 

Our findings will be of interest for applications such as coating and immersion lithography. Collisions between drops and the meniscus can occur in these geometries, but bubbles are usually not desired. We have shown that bubble sizes are strongly reduced when the initial contact occurs at a finite height above the substrate, which can be achieved by tuning the wettability of the substrate. However, we emphasize here that the geometry of impact can be much more complicated than the conditions studied in this paper. In our experiments we avoided strongly convex or concave menisci, such that the small sessile drops always impact a nearly flat meniscus. Meniscus curvature indeed has a strong influence, in particular on the drainage dynamics of the air film. As mentioned in Fig.~\ref{fig:chap4_overview}, the upward coalescence does not always lead to the entrapment of a bubble. We have not been able to quantify the parameters that determine whether or not a floating bubble will form. This remains an important open question, which needs to be answered to verify the validity of simple geometric arguments as a predictive tool.

\begin{figure}
	\centering 	
	\includegraphics[width=0.6 \textwidth]{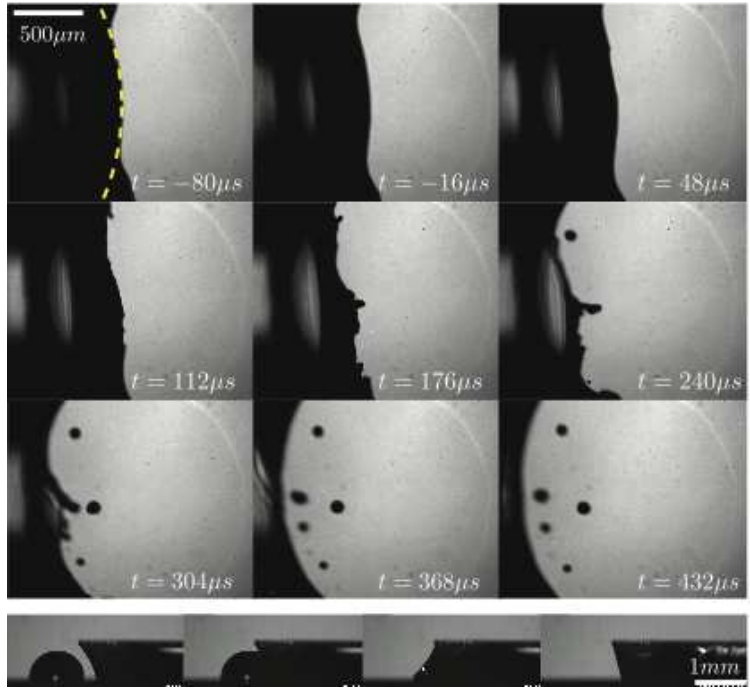}
		\caption{Complex break-up pattern of an air sheet (relatively large impacting droplet, $R = 1.1$ mm; height liquid bridge: $1.44$ mm).} 
	\label{fig:fig_chap4_discussion}
\end{figure}

The importance of geometry is illustrated in Fig.~\ref{fig:fig_chap4_discussion}, reporting the impact of a relatively large drop, with a drop size comparable to the height of the liquid bridge.  In that case, the shape of the air sheet and the subsequent drainage dynamics is strongly affected ($R$ becomes comparable with the radius of curvature of the meniscus). One can clearly observe the destabilization of the air sheet, and how this results into multiple small bubbles. This is very different from the impact of small drops, for which we always observed break-up of a single filament. We speculate that the relevant parameter is the drop size in relation to the relevant dimensions of the meniscus, such as its curvature or its height, but this remains to be investigated. Also, in the theoretical approach we have focused on the cases where the impact is normal to the contact line of the meniscus. An inclined impact, could of course have an effect on the symmetry and build up of the lubricating air film, and the subsequent coalescence process. It would be interesting to investigate these effects in more detail, both experimentally and numerically.

\section{Acknowledgments}
This work is part of the research programme 'Contact Line Control during Wetting and Dewetting' (CLC) of the 'Stichting voor Fundamenteel Onderzoek der Materie (FOM)', which is financially supported by the 'Nederlandse Organisatie voor Wetenschappelijk Onderzoek (NWO)'. The CLC programme is co‐financed by ASML and Océ.


\end{document}